\documentclass{an}
\usepackage{psfig}
\usepackage{times}
\Volume{000}             
\Year{0000}              
\Month{00}               
\Pagespan{000}{000}      

\begin{document}
\lhead[\thepage]{S.H. Saar \& A. Brandenburg: A New Look at Dynamo Cycle 
Amplitudes}
\rhead[Astron. Nachr./AN~{\bf XXX} (200X) X]{\thepage}
\headnote{Astron. Nachr./AN {\bf 32X} (200X) X, XXX--XXX}

\title{A New Look at Dynamo Cycle Amplitudes}

\author{Steven H. Saar\inst{1} and Axel Brandenburg\inst{2}}
\institute{Harvard-Smithsonian Center for Astrophysics, 60 Garden St.,
Cambridge, MA 02138, USA
\and
NORDITA, Blegdamsvej 17, DK-2100 Copenhagen \O, Denmark}
\date{Received {\it date will be inserted by the editor}; 
accepted {\it date will be inserted by the editor}} 

\abstract{We explore the dependence of the amplitude of stellar dynamo cycle
variability (as seen in
the Mount Wilson Ca {\sc II} HK timeseries data) on other stellar parameters.  
We find that the fractional cycle amplitude $A_{\rm cyc}$ 
(i.e.\ the ratio of the peak--to--peak variation to the average)
decreases somewhat with mean activity, increases with decreasing 
effective temperature, 
but is not correlated with inverse Rossby number $Ro^{-1}$. 
We find that $A_{\rm cyc}$ increases with the ratio of 
cycle and rotational frequencies $\omega_{\rm cyc}/\Omega$ along
two, nearly parallel branches.  
}
\correspondence{saar@head-cfa.harvard.edu}

\maketitle

\section{Introduction}

In a recent series of papers (Brandenburg, Saar, \& Turpin 
1998; Saar \& Brandenburg 1999 [=SB], 2001), we have been exploring
relationships between magnetic cycle periods $P_{\rm cyc}$,
rotation periods $P_{\rm rot}$, and other stellar properties. 
By combining $P_{\rm cyc}$ obtained from observations of Ca {\sc II} emission
(e.g., Baliunas et al. 1995 [=Bea95]),  photometry, and $P_{\rm rot}$
variation (e.g., Lanza \& Rodono 1999), we found evidence for
trends between cycle ($\omega_{\rm cyc}$) and rotational ($\Omega$)
frequencies (Saar \& Brandenburg 2001),  and between 
$\omega_{\rm cyc}/\Omega$ and both $R'_{\rm HK}$ and $Ro^{-1}$ 
(Brandenburg et al. 1998; SB).   Here, $Ro^{-1}$ is the inverse
Rossby number and $R'_{\rm HK}$ is the
Ca {\sc II} HK flux, corrected for photospheric contributions and
normalized by the bolometric flux (see Noyes et al. 1984).

Another important, but less studied observational property of
stellar dynamos is the cycle {\it amplitudes} (i.e., the mean amplitude
of the cyclic variability as seen in some activity diagnostic).
There is a relationship between cycle amplitude and period seen in the
Sun, where shorter cycles tend to be followed by stronger ones 
(e.g., Hathaway, Wilson \& Reichmann 1994), but there has 
been little work along these lines in stars.
Focusing on inactive stars, Soon et al. (1994) suggested that  
the fractional peak-to-peak cycle amplitude seen 
in chromospheric Ca {\sc II} HK emission,

\noindent $A_{\rm cyc} = \Delta R'_{\rm HK}/\langle R'_{\rm HK}\rangle$, 

\noindent decreased linearly with $\log (\Omega/\omega_{\rm cyc})^2$.  
Studying a larger sample, Baliunas et al. (1996) found that 
among inactive stars, $A_{\rm cyc} \propto (\omega_{\rm cyc}/\Omega)^{0.9}$.
Results for active stars were less clear.

Baliunas et al. (1996) analyzed {\it all} cycles detected by
Bea95, independent of their quality. 
In our earlier work on cycle periods we found
it very useful to begin by taking only a critically
chosen sample of reliable, high grade cycles (e.g.,
Bradenburg et al. 1998, SB). In this paper, we therefore 
take a fresh look at cycle amplitudes, in
light of the new results for $P_{\rm cyc}$,
using the same high quality cycle sample.

\section{Analysis}

We base our analysis on the Mount Wilson Ca {\sc II} HK data 
(Bea95).  Soon et al. (1994) gives cycle amplitudes for a number of 
inactive stars, and Baliunas et al. (1996) gives them
for all stars with cycles in Bea95, but the $A_{\rm cyc}$
values for common stars in these studies could not be reconciled. 
We have thus resorted, for this initial study,
to ``eye" estimates of the $S_{\rm HK}$ index amplitude $\Delta S_{\rm HK}$ 
(based on plots 
in Bea95) for stars not in Soon et al. (1994). These were converted into
a $\Delta R'_{\rm HK}$ following Noyes et al. (1984; see Table 1). 
As a test, we also made ``eye" estimates of $\Delta S_{\rm HK}$ 
for stars in Soon et al. (1994) and found
$\langle A_{\rm cyc}$(eye)$-A_{\rm cyc}$(Soon)$\rangle = 0.00 \pm 
0.03$, indicating good agreement.  

We studied stars chosen by SB for having
well defined cycles, which in practice meant all cycles ranked ``poor"
in Bea95, and some graded ``fair", were ignored. 
Stars were weighted by cycle grade, with an additional reduction 
applied if $A_{\rm cyc}$(eye) was unusually difficult to 
determine. Other stellar properties were also taken from the
list compiled in SB.  We estimated that the amplitude of the solar
secondary $P_{\rm cyc}$ (the Gleissberg cycle) at $\approx$ half
of the primary (11 year) $A_{\rm cyc}$. 

\begin{table}
\caption{Stellar Data}
\begin{tabular}{lccccc}\hline
HD \# & $B$-$V$ & $-$log & $A_{\rm cyc}^*$ & $P_{\rm rot}$ & $P_{\rm cyc}^*$ \\
      &       & $R'_{\rm HK}$ &  &  [d] & [yr] \\
\hline
Sun & 0.66    & 4.901 & 0.22 & 26.09 & 10.0,84 \\
1835 & 0.66 &  4.433 & 0.15 & 7.78 &  7.78   \\
3651 & 0.85   & 4.991 & 0.36 & 44.0 & 14.6 \\
4628 & 0.88   & 4.852 & 0.38 & 38.5 & 8.60 \\
10476 & 0.84  & 4.912 & 0.38 & 35.2 & 9.60 \\
16160 & 0.98  & 4.958 & 0.32 & 48.0 & 13.2 \\
18256 & 0.43 &  4.722 & 0.16   & 3.0 &  6.8 \\
20630 & 0.68 &  4.420 & 0.14 & 9.24 &  5.6    \\
26913 & 0.70 &  4.391 & 0.13 & 7.15 &  7.6    \\
26965 & 0.82  & 4.872 & 0.38 & 43.0 & 10.1 \\
32147 & 1.06  & 4.948 & 0.42 & 48.0 & 11.1 \\
76151 & 0.67 &  4.659 & 0.16 & 15. &  2.52    \\
78366 & 0.76 &  4.608 & 0.27,~0.19 & 9.67 &  12.2,~5.9 \\
100180 & 0.57 &  4.922 & 0.17,~0.07 & 14. &  12.9,~3.6 \\
103095 & 0.75 & 4.896 & 0.27 & 31.0 & 7.30 \\
114710 & 0.57 &  4.475 & 0.21,~0.12 & 12.35 &  16.6,~9.6 \\
115404 & 0.93 &  4.480 & 0.16 & 18.46 &  12.4     \\
136202 & 0.54 &  5.088 & 0.21 & 16. &  23.      \\
149661 & 0.82 &  4.583 & 0.35,~0.15 & 21.07 &  16.2,~4.0 \\
152391 & 0.61 &  4.448 & 0.28 & 11.43 &  10.9      \\
154417 & 0.57 &  4.533 & 0.17 & 7.78 &  7.4   \\
156026 &  1.16 &  4.662 & 0.37 & 21. &  21.       \\
160346 & 0.96 & 4.795 & 0.44 & 36.4 & 7.00 \\
165341 & 0.86 &  4.548 & 0.54,~0.12 & 19.9 &  15.5,~5.1 \\
166620 & 0.87 & 4.955 & 0.30 & 42.4 & 15.8 \\
187691 & 0.55 &  5.026 & 0.11 & 10. &  5.4      \\
190007 &  1.17 &  4.692 & 0.30 & 28.9 &  13.7      \\
190406 & 0.60 &  4.797 & 0.34,~0.15 & 13.94 &  16.9,~2.6 \\
201091 & 1.18 & 4.764 & 0.32 & 35.37 & 7.30 \\
201092 & 1.37 & 4.891 & 0.21 & 37.84 & 11.7 \\
219834B& 0.91 & 4.944 & 0.29 & 43.0 & 10.0 \\
\hline
\multicolumn{6}{l}{$^*$data for a secondary cycle listed second 
where appropriate}
\end{tabular}
\end{table}

\section{Results and Discussion}

We first investigated how $\Delta R'_{\rm HK}$ depends on 
$\langle R'_{\rm HK}\rangle$ itself (Fig. 1).
We find $\Delta R'_{\rm HK} \propto \langle R'_{\rm HK}\rangle^{0.77}$
($\sigma = 0.17$ dex; fitting the primary $P_{\rm cyc}$ only). 
This implies that $A_{\rm cyc} \propto
\langle R'_{\rm HK}\rangle^{-0.23}$ -- fractional cycle amplitudes
{\it decrease} somewhat with increasing activity.

Some of the 
scatter about the fit can be explained by an additional dependence on  
$B-V$ color: $\Delta R'_{\rm HK}$ is larger in K stars than G or F
at fixed $R'_{\rm HK}$.  This is seen more clearly by 
plotting $A_{\rm cyc}$ vs. $B-V$ color (Fig. 2), which shows an
steady increase in  $A_{\rm cyc}$ from F through G stars, until a maximum
is reached by the mid K stars ($B-V \sim 1.0$). The averages by
spectral type are 
$\langle A_{\rm cyc}$(F)$\rangle = 0.17$, $\langle A_{\rm cyc}$(G)$\rangle = 
0.22$, and $\langle A_{\rm cyc}$(K)$\rangle = 0.35$.  This behavior
may be the result of a dependence of cycle amplitude on 
fractional convection zone depth, up to some limiting value in mid-K stars.  

Some of the scatter in Fig. 1 is also intrinsic.  If stellar
cycles are like the Sun, activity will be restricted in latitude. 
Stars with different inclinations $i$ will then exhibit different
apparent $A_{\rm cyc}$ (Radick et al. 1998; Knaack et al. 2001).
Analysis of the models in Knaack et al. (2001), however, indicates that
77\% of the measured $A_{\rm cyc}$ should lie within $\pm$15\% of
the mean, with only 23\% (those with $i \leq 39^\circ$) 
will range from -15\% to -46\% of $\langle A_{\rm cyc}\rangle$. 
Furthermore, stars more active
than the Sun are expected to have a wider latitude distribution of
activity (e.g., Schrijver \& Title 2001), and thus should show less scatter due to varying $i$.
Thus the effects of $i$ should be relatively small, and strongest
stars with $R'_{\rm HK} \la R'_{\rm HK}(\odot)$.

\begin{figure}[ht]
\psfig{file=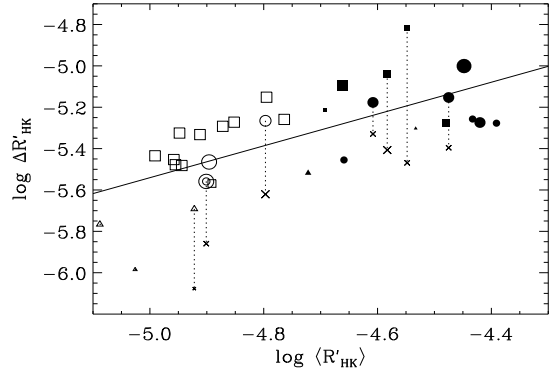,height=5.5cm,width=8cm}
\caption{$\Delta R'_{\rm HK}$ vs. $\langle R'_{\rm HK}\rangle$
for the stellar sample; symbols indicate 
F ($\triangle$), G ($\circ$, the Sun's is doubled), and K (box) 
stars, filled symbols are more active stars, size $\propto$ $P_{\rm cyc}$ and
$A_{\rm cyc}$ ``quality",  dotted lines connect secondary $P_{\rm cyc} (\times)$
in some stars.  We find $\Delta R'_{\rm HK} \propto \langle 
R'_{\rm HK}\rangle^{0.77}$ (solid line), implying $A_{\rm cyc}
\propto \langle R'_{\rm HK}\rangle^{-0.23}$.}
\label{figlabel}
\end{figure}

\begin{figure}[ht]
\psfig{file=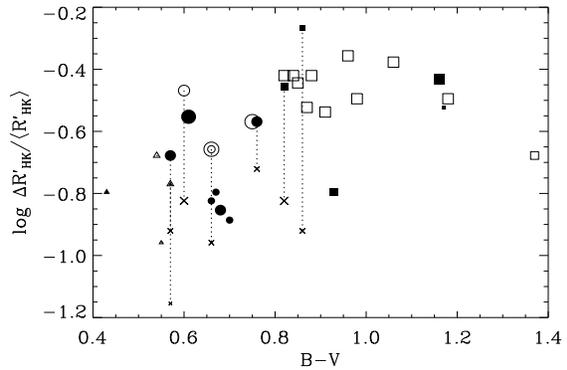,height=5.5cm,width=8cm}
\caption{$A_{\rm cyc} = \Delta R'_{\rm HK}/\langle R'_{\rm HK}\rangle$
vs. $B-V$ (symbols as in Fig. 1), showing an increasing $A_{\rm cyc}$
with decreasing effective temperature in F and G stars, reaching a maximum
in mid-K stars.}
\label{figlabel}
\end{figure}

In contrast to the dependence on $B-V$, $A_{\rm cyc}$ shows little
dependence on rotation, whether expressed as $\Omega$, or
(see Fig. 3) as $Ro^{-1} = \tau_C/P_{\rm rot}$
(where $\tau_C$ is the convective turnover timescale; Noyes et al. 1984).
Since $\langle R'_{\rm HK}\rangle$ shows a clear dependence on $Ro^{-1}$ 
(Noyes et al. 1984), this result implies $\Delta R'_{\rm HK}$
has a similar dependence on rotation.  Indeed, for $Ro^{-1} \leq 1.5$, 
$\Delta R'_{\rm HK} \propto Ro^{-1}$  (above this
$Ro^{-1}$, there is evidence for cycle amplitude saturation).

Since $A_{\rm cyc}< 1$, one might infer that only portion of 
$R'_{\rm HK}$ derives from a cycling dynamo.  This suggests that the 
non-cycling (small-scale) component of the dynamo is prominent
and thus (from Fig. 3) must have a 
similar dependence on rotation as the cycling component.
However, this is at odds with 
the lack of any strong rotational dependence for $R'_{\rm HK}$
in very inactive, ``flat" activity stars where activity is
likely fueled by a small-scale dynamo alone (Saar 1998).
Thus, a more likely possibility is that there is significant temporal 
overlap in $R'_{\rm HK}$ between cycles, reducing the
{\it apparent} amplitude $A_{\rm cyc}$.  In this scenario,
the cyclic dynamo dominates the rotational dependence of 
$R'_{\rm HK}$, despite the apparently small $A_{\rm cyc}$. 
Enhanced cycle overlap with increasing $\langle R'_{\rm HK}\rangle$
would also explain the decrease in $A_{\rm cyc}$
with $\langle R'_{\rm HK}\rangle$ (Fig. 1).

\begin{figure}[ht]
\psfig{file=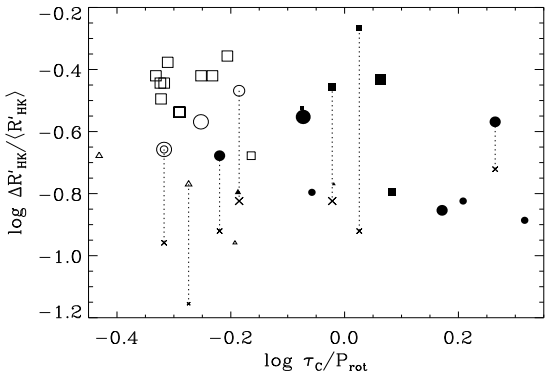,height=5.5cm,width=8cm}
\caption{$A_{\rm cyc} = \Delta R'_{\rm HK}/\langle\Delta R'_{\rm HK}\rangle$
vs. $Ro^{-1} = \tau_C/P_{\rm rot}$ (symbols as in Fig. 1), showing no
clear relationship.}  
\label{figlabel}
\end{figure}

Study of the dependence of $A_{\rm cyc}$ on $P_{\rm cyc}$
with our dataset reveals a complex situation.  Most inactive
stars (defined as $\log R'_{\rm HK} \leq -4.75$), 
combined with a few active ones,
trace out a relation similar to that found by Baliunas et al. (1996),
while most active ones show a similar, nearly parallel relation offset at
higher $A_{\rm cyc}$.
Specifically, 20 stars (14, or 70\% of them inactive) show
$A_{\rm cyc} \propto (\omega_{\rm cyc}/\Omega)^{0.66}$ ($\sigma = 0.082$ dex),
and 9 stars (7, or 78\% of them active) exhibit 
$A_{\rm cyc} \propto (\omega_{\rm cyc}/\Omega)^{0.85}$ ($\sigma = 0.081$ dex).
Only two lower quality detections cannot be assigned to one of these branches.
This ``branched" structure is similar to that seen 
between $\omega_{\rm cyc}/\Omega$ and $Ro^{-1}$ or $R'_{\rm HK}$ 
by Brandenburg et al. (1998),
SB, and Saar \& Brandenburg (2001) (and first noted by 
Saar \& Baliunas 1992 and Soon et al. 1993). 
There are some differences in branch membership, though;
for example, active G stars HD 1835, 20630, and 26913 lie on
the active (``A") branch in SB, but the inactive (``I") 
branch here (Figs. 4, 5).

\begin{figure}[ht]
\psfig{file=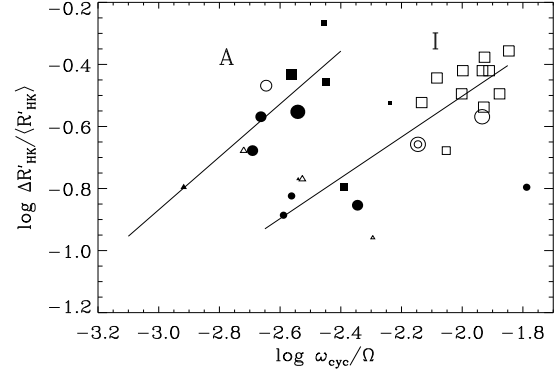,height=5.5cm,width=8cm}
\caption{$A_{\rm cyc} = \Delta R'_{\rm HK}/\langle\Delta R'_{\rm HK}\rangle$
vs. $\omega_{\rm cyc}/\Omega$ (symbols as in Fig. 1), showing an 
inactive (marked ``I"; $A_{\rm cyc} \propto (\omega_{\rm cyc}/\Omega)^{0.66}$) 
and an active (marked 
``A"; $A_{\rm cyc} \propto (\omega_{\rm cyc}/\Omega)^{0.85}$)
branch. }  
\label{figlabel}
\end{figure}

If we also consider secondary $P_{\rm cyc}$, we find three lie on 
these branches, while four do not.  Including the secondary $P_{\rm cyc}$
in the branch fits  causes little change: for the ``I" branch, 
$A_{\rm cyc} \propto (\omega_{\rm cyc}/\Omega)^{0.68}$ ($\sigma = 0.082$ dex)
while for the ``A" branch, 
$A_{\rm cyc} \propto (\omega_{\rm cyc}/\Omega)^{0.88}$ ($\sigma = 0.076$ dex).
Four of the stars show $A_{\rm cyc}$(primary)/$A_{\rm cyc}$(secondary) 
= 2.26$\pm$0.18, suggesting there may be some ``preferred" amplitude ratios.
All four secondary cycles ``unmatched"
to a branch lie at low $A_{\rm cyc}$ and high $\omega_{\rm cyc}/\Omega$ and
may indicate a third branch -- more data is needed to confirm this.

Thus, it appears that $A_{\rm cyc}$ and $P_{\rm cyc}$
are related in stars, just as they are in the Sun.
Study of stars indicates the relationship is multivalued, and depends on
rotation.

\begin{figure}
\psfig{file=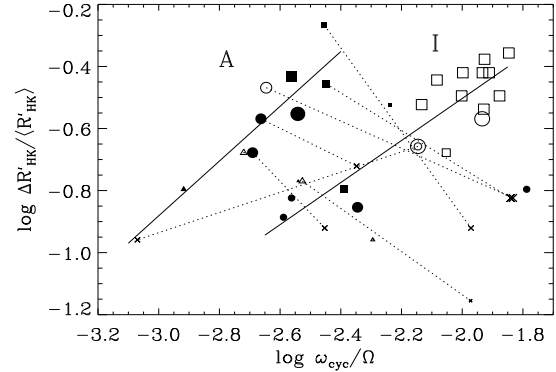,height=5.5cm,width=8cm}
\caption{Same as Fig. 4, but including secondary $P_{\rm cyc}$.
An inactive (marked ``I"; $A_{\rm cyc} \propto 
(\omega_{\rm cyc}/\Omega)^{0.68}$) and an active (marked ``A"; 
$A_{\rm cyc} \propto (\omega_{\rm cyc}/\Omega)^{0.88}$)
branch are indicated. }  
\label{figlabel}
\end{figure}

The present analysis is limited due to the saturation of Ca {\sc II} emission 
for $Ro^{-1} \ga 3$, which likely suppresses and 
obscures the visibility of cycles for more active stars.  Cycle overlap
is also a concern.
A logical next step would be to investigate photometric cycle amplitudes,
which in active stars are due to spots (rather than the plage/network
seen in Ca {\sc II}). Photometric cycle amplitudes saturate at
considerably higher $Ro^{-1}$ (Messina et al. 2001), permitting
study of $A_{\rm cyc}$ in much more active, faster rotating stars.  
Ideally, some normalized  quantity like the
fractional luminosity amplitude $\Delta L/L$ or the starspot
filling factor $f_S$ (e.g., from TiO measurements)
should be used.



%

\acknowledgements  This research was supported by NSF grant AST-9731652,
HST grant GO-8143, and NASA Origins grant NAG5-10630.
We are indebted to the referee, K. Ol\'ah for comments and suggestions, 
M. Dikpati for enlightening discussions, and to 
S. Baliunas and R. Donahue for their initial advice and encouragement.


\end{document}